\documentclass[aoas,MSNbibl,nameyear,dvips]{arximspdf}
\usepackage{graphicx}
\doi{10.1214/13-AOAS647} 
\volume{7}
\issue{3}
\pubyear{2013}
\firstpage{1525}
\lastpage{1539}



\begin{document}
\begin{frontmatter}

\title{Modeling and estimation of multi-source clustering in crime and
security data\thanksref{T1}}
\runtitle{Multi-source clustering in crime and security}
\thankstext{T1}{Supported in part by ARO Grant 58344-MA and NSF Grant
DMS-09-68309.}

\begin{aug}
\author{\fnms{George} \snm{Mohler}\corref{}\ead[label=e1]{gmohler@scu.edu}}
\runauthor{G. Mohler}
\affiliation{Santa Clara University}
\address{Department of Mathematics\\
\quad and Computer Science\\
Santa Clara University\\
713 Harding Ave\\
San Jose, California 95126\\
USA\\
\printead{e1}}
\end{aug}

\received{\smonth{8} \syear{2012}}
\revised{\smonth{2} \syear{2013}}

%
\begin{abstract}
While the presence of clustering in crime and security event data is
well established, the mechanism(s) by which clustering arises is not
fully understood. Both contagion models and history independent
correlation models are applied, but not simultaneously. In an attempt
to disentangle contagion from other types of correlation, we consider a
Hawkes process with background rate driven by a log Gaussian Cox
process. Our inference methodology is an efficient Metropolis adjusted
Langevin algorithm for filtering of the intensity and estimation of the
model parameters. We apply the methodology to property and violent
crime data from Chicago, terrorist attack data from Northern Ireland
and Israel, and civilian casualty data from Iraq. For each data set we
quantify the uncertainty in the levels of contagion vs. history
independent correlation.
\end{abstract}

%

\end{frontmatter}

\section{Introduction}\label{sec1}
\begin{figure}

\includegraphics{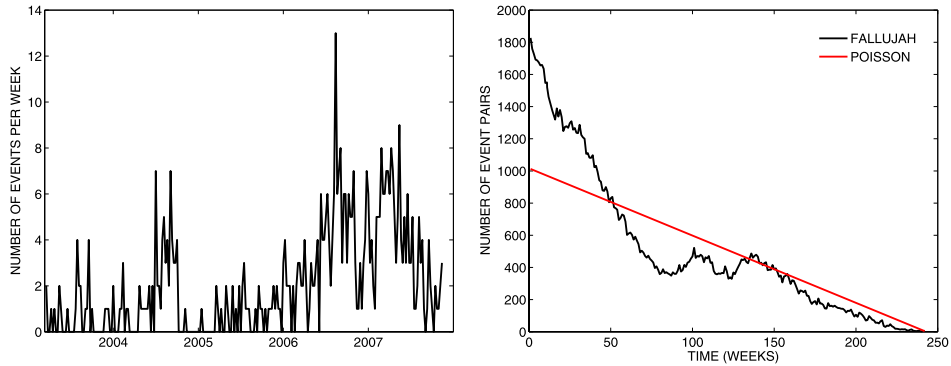}

\caption{Left: Civilian fatalities per week in Fallujah. Right:
Distribution of inter-event times.}\label{fig1}
\end{figure}

Self-exciting point processes have gained attention in recent years for
the purpose of modeling criminal activity, in particular, property
crime and gang violence [\citet
{Egesdal2010,hegemann2012,mohler2011}, \citeauthor{short2009} (\citeyear{short2009,short2010}), \citet{stomakhin2011}],
and, more recently, terrorism and other event patterns in extreme
security settings [\citet{lewis2012,porter2012}]. The defining
characteristic of these models is that the occurrence of an event
increases the likelihood of more events, as the offender(s) may attempt
to replicate a previous success in the same or a nearby location in the
following days or weeks [\citet{bowers2004,short2009,Townsley2008}]. In
\citet{short2009}, a simple procedure is introduced to detect
self-excitation in event data, where the distribution of inter-event
times $t_i-t_j$ for all $i>j$ is compared to the theoretical
distribution corresponding to a stationary Poisson process. For
example, we plot in Figure~\ref{fig1} a histogram of the inter-event times
$t_i-t_j$ ($i>j$) for civilian casualties per week in Fallujah between
March 20, 2003 and December~31, 2007 provided by Iraq Body Count (IBC).
The histogram is an estimate of the unnormalized density of inter-event
times and is similar to the $K$-function estimator in \citet{moller2003}
[for a uniform distribution on an interval the function decreases
linearly, see \citet{short2009} for further details]. The presence of
more event pairs at shorter inter-event times compared to random chance
provides some evidence that self-excitation may play a role in the
occurrence of fatal attacks in Fallujah.

However, other second order point processes are capable of producing
clustered inter-event times as depicted in Figure~\ref{fig1}. In particular,
auto regressive and Cox processes have also been employed as potential
models for crime and security related event clustering [\citet
{taddy2010,zammit2012}]. In these models events are correlated through
the intensity of the process, which follows a random trajectory.
Whereas for a self-exciting point process the intensity will stay high
for a period of time following an event, for a Cox process the
intensity may quickly decrease following an event due to random
fluctuations. From a social perspective, events may be correlated due
to exogenous factors like the state of the economy, month of the year,
change in military operations, etc., rather than ``caused'' by
endogenous factors such as repeat offender behavior.

We propose a model along with an efficient inference methodology for
quantifying uncertainty in the levels of contagion vs. history
independent correlation in crime and security data sets. The model
consists of a discrete time Hawkes process with background rate
determined by a log Gaussian Cox process (LGCP). The Gaussian process
is given by the forward Euler discretization of a mean reverting
Ornstein--Uhlenbeck stochastic differential equation. Details of the
model are provided in Section~\ref{sec2}. For filtering of the intensity and
estimation of the model parameters, we consider an extension of the
Metropolis adjusted Langevin algorithm (MALA) for LGCPs to the case of
self-excitation. By exploiting properties of the covariance matrix of
the model in Section~\ref{sec2}, MALA can be implemented such that the cost of
each metropolis iteration scales linearly with the size of the data.
Details of the inference methodology are provided in Section~\ref{sec3}. In
Section~\ref{sec4} we validate the methodology on synthetic data and then apply
it to several open source crime and security data sets: property and
violent crime in Chicago, terrorist attacks in Northern Ireland and
Israel, and civilian casualties in Fallujah, Iraq. We confirm previous
work suggesting that contagion plays a role in crime event clustering,
though our results indicate that correlated fluctuations are also
important. For data sets corresponding to more extreme security
settings, we observe a wider range in the levels of
contagion.\looseness=1

\section{A Hawkes--Cox process model of crime and security}\label{sec2}

\begin{figure}

\includegraphics{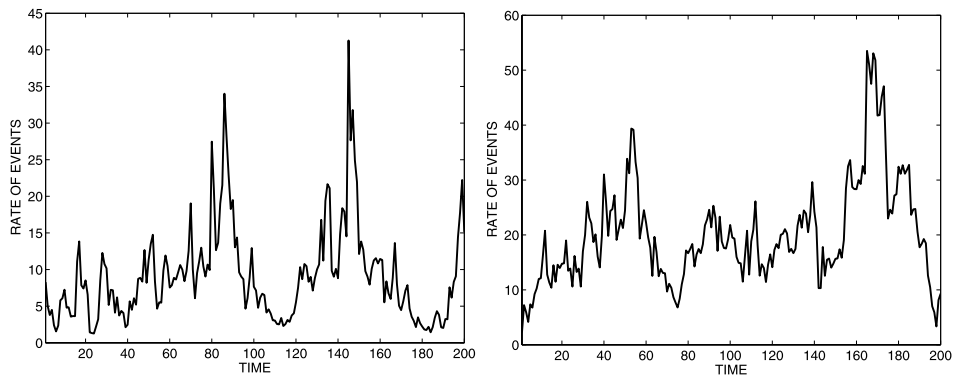}

\caption{Left: Cox process with parameters $a=0.9$, $\sigma^2=0.7$ and
$\mu=1.8$. Right: Hawkes process with parameters $\mu=0.8$, $b=0.075$ and
$\theta=0.9$.}\label{fig2}
\end{figure}

We consider a discrete time model for the intensity of events where the
background rate is determined by a Log Gaussian Cox process (LGCP) and
the intensity is self-excited by the occurrence of events. In
particular, the intensity is given by
%
%
\begin{equation}
\lambda_i=e^{x_i}+\sum_{i>j}
\theta\frac{(1-b)}{b} b^{i-j} y_j,\label{CH}
\end{equation}
where $y_i$ is the number of events and $\lambda_i$ is the expected
number of events in the time interval $[i\Delta t,(i+1)\Delta t]$. The
parameters $\theta$ and $b$ control the level and timescale of
contagion effects and we use the initial conditions $\lambda_0=e^{\mu}$
throughout. Here ${x_i}$ is a Gaussian process with mean $\mu$ and
covariance matrix $\Sigma$, where
%
%
\begin{equation}
\Sigma_{ij}=\sigma^2 a^{|i-j|}.
\end{equation}
The model is capable of producing event clustering due to both
contagion effects and history independent correlations. For example,
$\theta=0$ corresponds to a LGCP and $\sigma^2=0$ corresponds to a
discrete time version of a Hawkes process. The parameters $a$ and $b$
control the timescales over which history independent correlation and
contagion effects persist. In Figure~\ref{fig2} we plot two realizations of the
intensity~(\ref{CH}), one corresponding to a LGCP without
self-excitation and one corresponding to a Hawkes process with constant
background rate. We note that in both cases significant clustering is
observed and it is difficult to distinguish the type of clustering
based upon visual inspection of the intensity. We will return to this
example in Section~\ref{sec4}.

%

The model given by equation (\ref{CH}) is related to the standard
continuous time Hawkes process,
%
%
\begin{equation}
\lambda(t)=\nu+\sum_{t>t_i}k(t-t_i),
\label{Hawkes}
\end{equation}
where $\lambda$ is the intensity (rate) of events, $\nu$ is the
background (Poisson) rate of events, and $k(t-t_i)$ is a triggering
kernel that determines the distribution of offspring events generated
by events $t_i$ from the history of the process. Hawkes processes are
used to model risk increases triggered by events in the history of the
process and have been applied to repeat offender behavior in burglary
[\citet{mohler2011}], retaliations in gang violence [\citet
{Egesdal2010,hegemann2012,short2010,stomakhin2011}] and terrorist and
insurgent activity [\citet{lewis2011,porter2012}]. Continuous time Log
Gaussian Cox processes are also used to model event clustering, for
example, in \citet{brix2001} a mean reverting Ornstein--Uhlenbeck SDE is
used to determine the intensity of the point process. For an
exponential kernel (\ref{Hawkes}) can be written as a differential
equation and a continuous time Hawkes--Cox process is governed by the
system of stochastic differential equations,
%
%
\begin{eqnarray}
{dx_t}&=&-\omega_1(x_t-\mu)\,dt+
\alpha_1 \,dB_t, \label{sde1}
\\
{dg_t}&=&-\omega_2g_t \,dt+
\alpha_2 \omega_2 \,dN_t,k,\label{sde2}
\end{eqnarray}
where $B_t$ is a standard Brownian motion and $N_t$ is a point process
with conditional intensity
%
%
\begin{equation}
\lambda_t=e^{x_t}+g_t.
\end{equation}
Here the parameters $\omega_1$ and $\omega_2$ determine the timescale
over which clustering due to history independent correlations and
contagion last, $\alpha_1$ and $\alpha_2$ determine the size of
intensity fluctuations, and $\mu$ determines the baseline level of
event activity.

To facilitate simulation and estimation, in addition to the fact that
many crime and security data sets are binned by day or some other time
unit, we restrict our attention to discrete forward Euler approximations,
%
%
\begin{eqnarray}
{x_i}&=&x_{i-1}-\omega_1(x_{i-1}-\mu)
\Delta t+\alpha_1 \sqrt{\Delta t} Z_{i-1},
\\
{g_i}&=&g_{i-1}-\omega_2g_{i-1} \Delta
t+\alpha_2 \omega_2 y_{i-1},
\end{eqnarray}
where $Z_{i-1}=\mathcal{N}(0,1)$ and $y_{i-1}$ is the number of events
occurring in $[(i-1)\Delta t,i\Delta t]$.
Letting $a=(1-\omega_1 \Delta t)$, $\sigma^2=\alpha_1^2 \Delta
t/(1-(1-\omega_1\Delta t)^2)$, $b=(1-\omega_2\Delta t)$ and $\theta
(1-b)=\alpha_2 \omega_2$, the discrete model takes the form of (\ref{CH}).

\section{Filtering and estimation using MALA}\label{sec3}
In this section we discuss general strategies for point process
estimation and then develop an inference methodology for simultaneously
filtering the latent Gaussian process $x_i$ and estimating the
parameters $(a,\sigma^2,\mu,b,\theta)$ from observations $y_i$ assumed
to have been generated by a model of the form (\ref{CH}). Our goal will
be to quantify the uncertainty in the level of contagion present in an
event time series, as well as to detect history independent correlation
that may also be present. In particular, we will use Markov Chain Monte
Carlo to obtain a posterior probability distribution for the intensity
of the process, the latent Gaussian process and the model parameters.

For Hawkes processes with stationary background rate, EM-type
algorithms are a popular choice and parametric and variational versions
are used in seismological and security related applications [\citet
{lewis2011,marsan2008,mohler2011,Sornette2009,Veen2008}]. One issue
that arises, however, is that parameter estimates suffer from high
variance when the clusters are not well separated [\citet
{lewis2011,Sornette2009}], a common problem for the EM algorithm
applied to mixture models (the Hawkes process in equation \ref{Hawkes}
can be viewed as a mixture model with the number of mixtures equal to
the number of points in the data set). With the introduction of a
nonstationary LGCP background rate, EM estimates are likely to have
even higher variation. We will therefore take a Bayesian approach to
the estimation problem in order to quantify the uncertainty in
parameter estimates.

We note that a variety of methods have been developed for the
estimation of temporal point processes, including EM algorithms [\citet
{smith2003}], variational alternatives [\citet{mangion2011}],
integrated nested Laplace approximations [\citet{rue2009}] and
expectation propagation [\citet{cseke2011}]. As discussed in \citet
{brix2001}, sequential filtering for LGCPs suffers from large variance
of the importance weights and the authors instead use the Metropolis
adjusted Langvien algorithm (MALA) for filtering the intensity of LGCPs
after estimating the parameters via a moment-based method. Similar jump
diffusion models to (\ref{sde1})--(\ref{sde2}) have recently been used
to model financial contagion. In \citet{giesecke2011}, an approximate
likelihood filter is employed that avoids the need for Monte Carlo
simulation, though the computational cost of the method prevents the
straightforward extension to spatial processes. For simultaneous
filtering of the intensity and estimation of parameters, Langevin and
Hamiltonian Monte Carlo methods on manifolds are developed in \citet
{girolami2011} capable of handling high-dimensional/spatial problems.
We take this approach as well, though we avoid the need for manifold
based Monte Carlo by exploiting an analytic expression for the inverse
covariance matrix of the process.

\subsection{Metropolis adjusted Langevin algorithm}\label{sec3.1}

In general,\vspace*{1pt} given a random vector $\vec{\theta}$ with density $\pi(\vec
{\theta})$, the stochastic differential equation (Langevin equation),
%
%
\begin{equation}
d\vec{\theta}(t)=\nabla_{\vec{\theta}} \log \bigl(\pi(\vec{\theta})
\bigr)\,dt/2+dB(t),\label{Langevin}
\end{equation}
has stationary distribution $\pi(\vec{\theta})$. The forward Euler
discretization of (\ref{Langevin}) is given by
%
%
\begin{equation}
\vec{\theta}^*=\vec{\theta}^n+\frac{\varepsilon^2}{2}\nabla_{\vec{\theta}}
\log \bigl(\pi\bigl(\vec{\theta}^n\bigr) \bigr)+\varepsilon
Z_n,\label{Euler}
\end{equation}
which no longer has the correct stationary distribution, nor satisfies
detailed balance. These shortfalls are overcome through MALA by
adjusting the Langevin equation with a Metropolis acceptance condition
after each Euler step. The transition density is given by
%
%
\begin{equation}
q\bigl(\vec{\theta}^*|\vec{\theta}^n\bigr)=\mathcal{N} \biggl(\vec{
\theta}^n+\frac
{\varepsilon^2}{2}\nabla_{\vec{\theta}} \log \bigl(\pi\bigl(
\vec{\theta}^n\bigr) \bigr),\varepsilon^2 I \biggr)
\label{transition}
\end{equation}
and the acceptance probability is
%
%
\begin{equation}
\min \biggl\{1,\frac{q(\vec{\theta}^n|\vec{\theta}^*)\pi(\vec{\theta
}^*)}{q(\vec{\theta}^*|\vec{\theta}^n)\pi(\vec{\theta}^n)} \biggr\}.\label {proposal}
\end{equation}
The posterior density for the discrete Hawkes--Cox process is given by
%
%
\begin{eqnarray}\label{likelihood}
&&\pi\bigl(\vec{x},a,\sigma^2,\mu,b,\theta|\vec{y}\bigr)\nonumber\\
&&\qquad\propto
 \Biggl(\prod_{i=1}^N \exp\{-
\lambda_i\}\lambda_i^{y_i} \Biggr)
 |\Sigma|^{-1/2}\exp\bigl\{-(\vec{x}-\mu1)^T
\Sigma^{-1}(\vec{x}-\mu1)/2\bigr\}\\
&&\qquad\quad{}\times p\bigl(a,\sigma^2,\mu,b,\theta\bigr),\nonumber
\end{eqnarray}
where $p(a,\sigma^2,\mu,b,\theta)$ is the prior distribution of the
model parameters. The derivatives of the posterior density are given by
%
%
\begin{equation}
\nabla_{\vec{x}}\log(\pi)=\vec{v}-\Sigma^{-1}(\vec{x}-\mu1),
\label{Dx}
\end{equation}
where $v_i=y_i \exp\{x_i\}/\lambda_i-\exp\{x_i\}$,
%
%
\begin{eqnarray}\label{Da}
\nabla_{a}\log(\pi)&=&-0.5\frac{d\log(|\Sigma|)}{da}+0.5(\vec{x}-\mu
1)^T\Sigma^{-1}\frac{d\Sigma}{da}\Sigma^{-1}(
\vec{x}-\mu1)
\nonumber
\\[-8pt]
\\[-8pt]
\nonumber
&&{}+\frac{d\log
(p)}{da},
\\
\label{Dsigma}\nabla_{\sigma^2}\log(\pi)&=&-0.5\frac{d\log(|\Sigma|)}{d\sigma^2}+0.5(\vec {x}-
\mu1)^T\Sigma^{-1}\frac{d\Sigma}{d\sigma^2}\Sigma^{-1}(
\vec{x}-\mu 1)
\nonumber
\\[-8pt]
\\[-8pt]
\nonumber
&&{}+\frac{d\log(p)}{d\sigma^2},
\\
\nabla_{\mu}\log(\pi)&=&\sum_{i=1}^N
\bigl(\Sigma^{-1}(\vec{x}-\mu1) \bigr)_i+
\frac{d\log(p)}{d\mu},\label{Dmu}
\\
\nabla_{b}\log(\pi)&=&\sum_{i=1}^N
(y_i/\lambda_i-1)\frac{d\lambda
_i}{db}+\frac{d\log(p)}{db}
\label{Db}
\end{eqnarray}
and
%
%
\begin{equation}
\nabla_{\theta}\log(\pi)=\sum_{i=1}^N
(y_i/\lambda_i-1)\frac{d\lambda
_i}{d\theta}+\frac{d\log(p)}{d\theta}.
\label{Dtheta}
\end{equation}

In general, each Metropolis proposal is associated with $O(N^3)$
operations, as the inverse covariance matrix and the determinant are
required. However, these can be determined analytically for our
covariance matrix [\citet{shaman1969}]:
%
%
\begin{equation}
\Sigma^{-1}=\frac{1}{\sigma^2(1-a^2)}\left[ %
\matrix{ 1 & -a & 0 & \cdots& 0
\vspace*{2pt}\cr
-a & 1+a^2 & -a & \cdots& 0
\vspace*{2pt}\cr
0 & \ddots& \ddots& \ddots& 0
\vspace*{2pt}\cr
\vdots & & -a & 1+a^2 & -a
\vspace*{2pt}\cr
0 & \cdots& 0 & -a & 1}
\right]\label{inverse}
\end{equation}
and
%
%
\begin{equation}
|\Sigma|=\sigma^{2N}\bigl(1-a^2\bigr)^{N-1}.
\label{det}
\end{equation}
Because $\Sigma^{-1}$ is tridiagonal, (\ref{likelihood}) and (\ref{Dx})
require $O(N)$ operations to evaluate. Furthermore, the derivatives on
the right side of (\ref{Da}) and (\ref{Dsigma}) can be\vadjust{\eject} computed
directly from (\ref{likelihood}), (\ref{inverse}) and (\ref{det}). The
recursive relationship,
%
%
\begin{equation}
\lambda_i-e^{x_i}=b\bigl(\lambda_{i-1}-e^{x_{i-1}}
\bigr)+\theta(1-b)y_{i-1},
\end{equation}
can be used to compute $\lambda_i$, $\frac{d\lambda_i}{db}$ and $\frac
{d\lambda_i}{d\theta}$ efficiently in $O(N)$ operations.

\section{Results}\label{sec4}

In this section we validate the inference methodology of Section~\ref{sec3} on
synthetic data generated by the discrete Hawkes--Cox process model
introduced in Section~\ref{sec2}. We then apply the methodology to several open
source crime and terrorism data sets to estimate the levels of
contagion and history independent correlation present in the data.

For all examples we use the following MCMC iteration procedure. At each
iteration we alternately sample first the latent variable, $\vec{x}$,
using equations (\ref{transition})--(\ref{proposal}) with Langevin step
size $\varepsilon=0.1$, second the variables $a$, $\sigma^2$ and $\mu$ with
step size $\varepsilon=0.01$, and third the parameters $b$ and $\theta$
with step size $\varepsilon=0.01$. We note that the second and third steps
are independent, as the parameters are only coupled through their
dependence on $\vec{x}$. We use ${U}[0,1]$ priors for the parameters
$a$, $b$ and~$\theta$, and $\mathcal{N}(0,5)$ priors for $\mu$ and
$\sigma^2$ ($\sigma^2$ restricted to be positive). $5\cdot10^5$ Monte
Carlo iterations are used in each example with a burn-in of $2.5 \cdot
10^5$. Trace plots of the posterior distribution are inspected to
verify convergence.

\begin{figure}[b]

\includegraphics{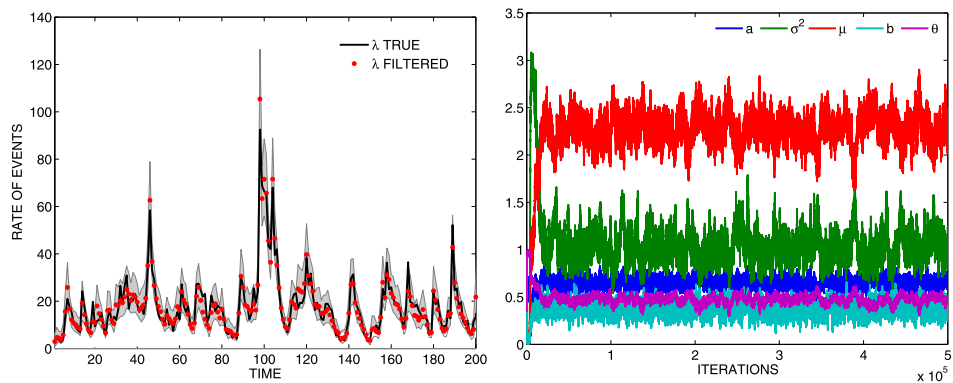}

\caption{Left: First 200 time units of the simulated intensity (black)
along with the filtered intensity (posterior mean in red, 95\% range in
grey) with $N=500$, $a=0.65$, $\sigma^2=1$, $\mu=2$, $b=0.35$, and $\theta
=0.5$. Right: Trace plots of the posterior parameter distributions.}\label{fig3}
\end{figure}

\subsection{Example 1: Two sources of correlation}\label{sec4.1}

We first validate the inference methodology for a discrete Hawkes--Cox
process with $N=500$ time steps and parameters $a=0.65$, $\sigma^2=1$,
$\mu=2$, $b=0.35$, and $\theta=0.5$. Convergence to the stationary
distribution is reached in less than $10^5$ MCMC iterations, as
illustrated by the trace plots of the posterior parameter distributions
shown in Figure~\ref{fig3}. For an arbitrary covariance matrix the cost of one
Monte Carlo step would be $O(500^3)$, but due to the linear dependence
on the size of the data, we were able to take $5\cdot10^5$ Monte Carlo
steps implemented in Matlab with a $1.8$~GHz dual-core Intel i7
processor in less than $2$ hours.

\begin{figure}

\includegraphics{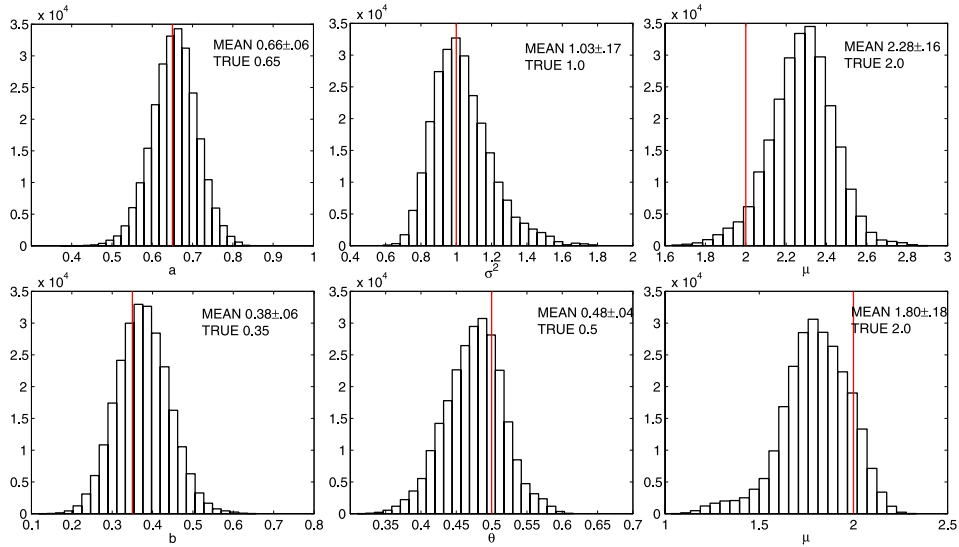}

\caption{Histograms of the sampled posterior parameter distributions
(250,000 samples) corresponding to a simulated point process with
parameters $a=0.65$, $\sigma^2=1$, $\mu=2$, $b=0.35$, and $\theta=0.5$.
The posterior mean and standard deviation are displayed in the top
right of each figure and the red line indicates the true value of the
parameter used to simulate the intensity. The lower right posterior
distribution for the parameter $\mu$ corresponds to a different
realization of the point process.}\label{fig4}\vspace*{-2pt}
\end{figure}

We confirm the accuracy of the methodology by plotting the filtered
intensity (posterior mean) against the true intensity of the simulated
Hawkes--Cox process (first 200 time units for visualization) in Figure~\ref{fig3}. The shaded region indicates the 95\% range of the posterior
intensity. In Figure~\ref{fig4} we plot the posterior distribution of the five
model parameters along with the true parameters (indicated by a red
vertical line) used in the simulation. The relative error between the
true parameter value and the posterior mean is less than $14\%$ for all
parameters. We note that for a particular realization of the point
process the posterior distribution for the estimate of $\mu$ appears
biased, however, for different realizations of the intensity the
estimate may over- or underestimate $\mu$.


\subsection{Example 2: Contagion vs. history independent correlation}\label{sec4.2}

We return to the example plotted in Figure~\ref{fig2} in order to assess
whether the Langevin Monte Carlo method can distinguish between a
Hawkes process and a Cox process.

\begin{figure}

\includegraphics{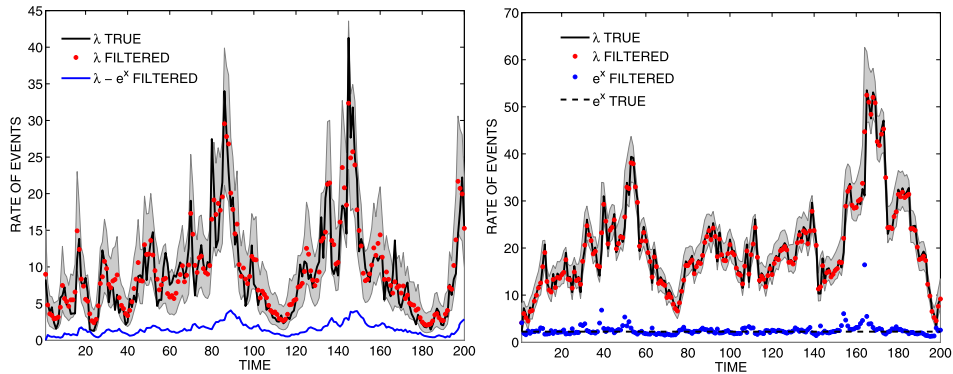}

\caption{Left: Simulated intensity (black) of a Cox process with
parameters $a=0.9$, $\sigma^2=0.7$ and $\mu=1.8$ along with the filtered
intensity (posterior mean in red, 95\% range in grey, and estimated
Hawkes portion of the intensity in blue). Right: Simulated intensity
(black) of a Hawkes process with parameters $\mu=0.8$, $b=0.075$ and
$\theta=0.9$ along with the filtered intensity (posterior mean in red,
95\% range in grey, and estimated Cox portion of the intensity in blue).}\label{fig5}
\end{figure}

For the Cox process we use the parameters $a=0.9$, $\sigma^2=0.7$ and
$\mu=1.8$ ($\theta=0$) with $N=200$. The Hawkes portion of the
intensity $\lambda-e^x$ thus equals zero in equation (\ref{CH}). In
Figure~\ref{fig5} (left) we plot the true intensity of the simulated Cox process
(black) along with the filtered intensity (posterior mean in red) and
the filtered Hawkes portion of the intensity (posterior mean in blue).
We note that the Hawkes portion of the estimated intensity remains
close to zero throughout the time interval, though for periods of high
event activity it accounts for up to $1/6$ of the overall rate of
events. Thus, one needs to be cautious in interpreting results for
similar levels of contagion inferred from crime and security data sets.

For the Hawkes process with constant background rate we use the
parameters $\mu=0.8$, $b=0.075$ and $\theta=0.9$ ($\sigma=0$) with
$N=200$. In Figure~\ref{fig5} (right) we plot the true intensity $\lambda$
(black) and background rate $e^x$ (dashed black) against the filtered
intensity (posterior mean in red) and Cox contribution to the estimated
intensity (blue dots). We note that the filtered background rate
exhibits low variation and provides a good approximation to the actual
background rate, $\exp(0.8)$.

\subsection{Application to crime and security data}\label{sec4.3}
Next we apply the inference methodology to several open source crime
and security data sets to assess the sources of clustering. The first
data set is the counts per week of property crime (burglary and motor
vehicle theft) and violent crime (battery, assault and robbery)
occurring in Beat 423 in Chicago between January 1, 2001 and June 15,
2012. The data is available through the Chicago data portal at
\texttt{\href{https://data.cityofchicago.org/Public-Safety/Crimes-2001-to-present/ijzp-q8t2}{https://data.cityofchicago.org/Public-Safety/Crimes-2001-to-}
\href{https://data.cityofchicago.org/Public-Safety/Crimes-2001-to-present/ijzp-q8t2}{present/ijzp-q8t2}}.
The terrorism data sets we use include the counts per week of attacks
in Israel (2001--2010) and Northern Ireland (1970--1993). The data is
available through the Global Terrorism Database at
\url{http://www.start.umd.edu/gtd/}. The civilian casualty data from
Fallujah described in Section~\ref{sec1} can be obtained through the IBC at
\texttt{\href{http://www.iraqbodycount.org/database/}{http://www.}
\href{http://www.iraqbodycount.org/database/}{iraqbodycount.org/database/}}.

\begin{figure}

\includegraphics{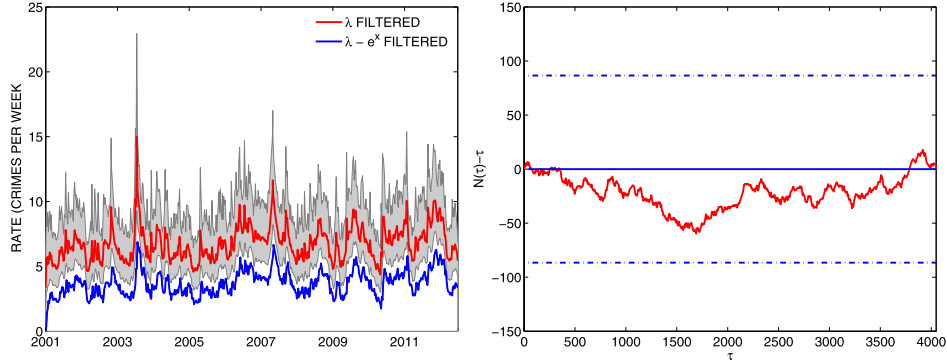}

\caption{Property crime in Chicago. Left: Filtered intensity $\lambda$
(posterior mean in red, 95\% range in grey) and Hawkes rate $\lambda
-e^x$ (posterior mean in blue). Right: Normalized cumulative
distribution of rescaled event times along with $95\%$ error bounds of
the Kolmogorov--Smirnov statistic.}\label{fig6}
\end{figure}

\begin{figure}[b]

\includegraphics{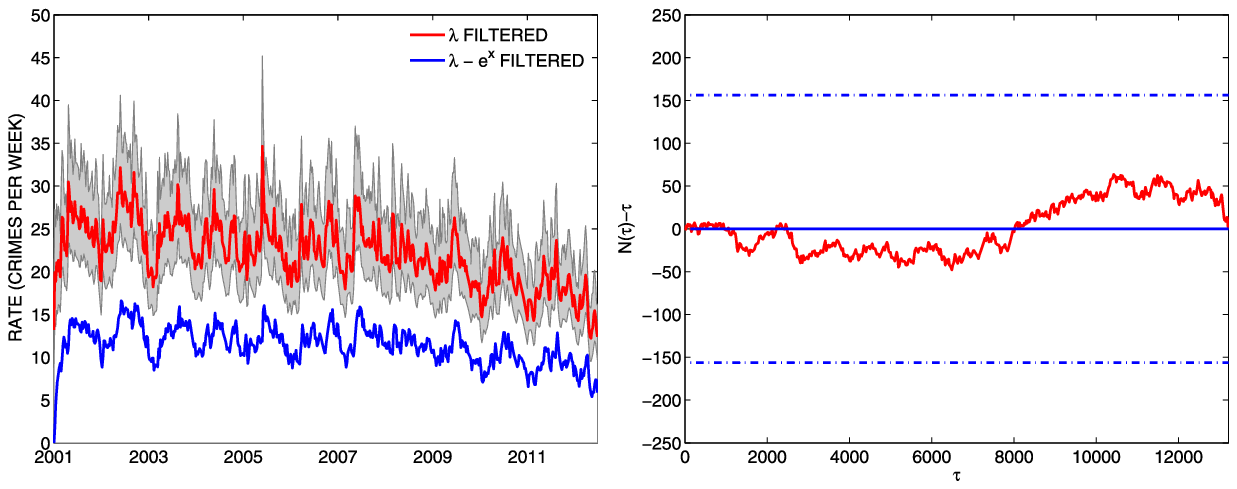}

\caption{Violent crime in Chicago. Left: Filtered intensity $\lambda$
(posterior mean in red, 95\% range in grey) and Hawkes rate $\lambda_i
-e^{x_i+c\cdot i}$ (posterior mean in blue). Right: Normalized
cumulative distribution of rescaled event times along with $95\%$ error
bounds of the Kolmogorov--Smirnov statistic.}\label{fig7}
\end{figure}

In Figure~\ref{fig6} we plot the filtered intensity for property crime in
Chicago and in Figure~\ref{fig7} we plot the filtered intensity for violent
crime. In order to assess the goodness of fit of the model, we use
residual analysis [\citet{ogata1988}]; the rescaled event times
%
%
\begin{equation}
\tau_i=\int_0^{t_i}\lambda(t)\,dt
\end{equation}
are distributed according to a unit rate Poisson process if the model
is correctly specified. On the right of Figures~\ref{fig6} and \ref{fig7} we plot the
normalized cumulative number of events $N(\tau)-\tau$ against the
rescaled event times $\tau$.

For property crime $45\%$ of the events are attributed to the
background rate $e^x$ and the other $55\%$ are attributed to the Hawkes
component (see Table~\ref{tab1}). The posterior standard deviation for the
percentage of events assigned to the Hawkes intensity in $7\%$, thus
indicating with a high degree of certainty that both types of
correlation are playing a significant role in intensity fluctuations.
The time scale parameter $\omega_1^{-1}$ of mean reversion of the Cox
process is estimated to be $1.7$ weeks and for the Hawkes triggering
kernel the time scale $\omega_2^{-1}$ is estimated to be $4.8$ weeks. A
number of exogenous factors fluctuating on a weekly basis could be
causing the correlated fluctuations, such as weather or routine
activities linked to work and pay schedules. The several week duration
of self-excitation is consistent with previous estimates for property
crime. We also note that the normalized cumulative distribution of
rescaled event times stays well within the $95\%$ error bounds of the
Kolmogorov--Smirnov statistic.

\begin{table}
\caption{Posterior mean and standard deviation of the percentage of
events attributed to the Hawkes component of the estimated intensity
(top row) and posterior mean of the timescales $\omega_1^{-1}$ and
$\omega_2^{-1}$ in weeks associated with the Cox and Hawkes processes,
respectively (bottom two rows)}\label{tab1}
\begin{tabular*}{\textwidth}{@{\extracolsep{\fill}}lccccc@{}}
\hline
& \textbf{Prop.} & \textbf{Viol.} & \textbf{N. Ireland} & \textbf{Israel} & \textbf{Fallujah} \\
\hline
{\% Hawkes} & {55} (7) & {52} (7) & {50} (5) & 12 (7) & 23 (13) \\
{Timescale Cox} & {1.7} & {1.5} & {1.7} & {2.8} & 36.0 \\
{Timescale Hawkes} & {4.8} & {3.9} & {9.3} & {5.7} & \phantom{0}4.9 \\
\hline
\end{tabular*}
\end{table}

The Chicago violent crime data and the Northern Ireland terrorist
attack data both exhibit significant slow timescale trends over the
observation window. To account for this in the model, we multiply the
background rate $e^{x_i}$ by an exponential factor $e^{c\cdot i}$ and
estimate $c$ along with the other model parameters using MALA. For
violent crime we observe similar levels of contagion and correlation,
as well as similar timescales to the property crime time series. With
the addition of the exponential factor in the model, $N(\tau)-\tau$
stays well within the $95\%$ error bounds.

\begin{figure}

\includegraphics{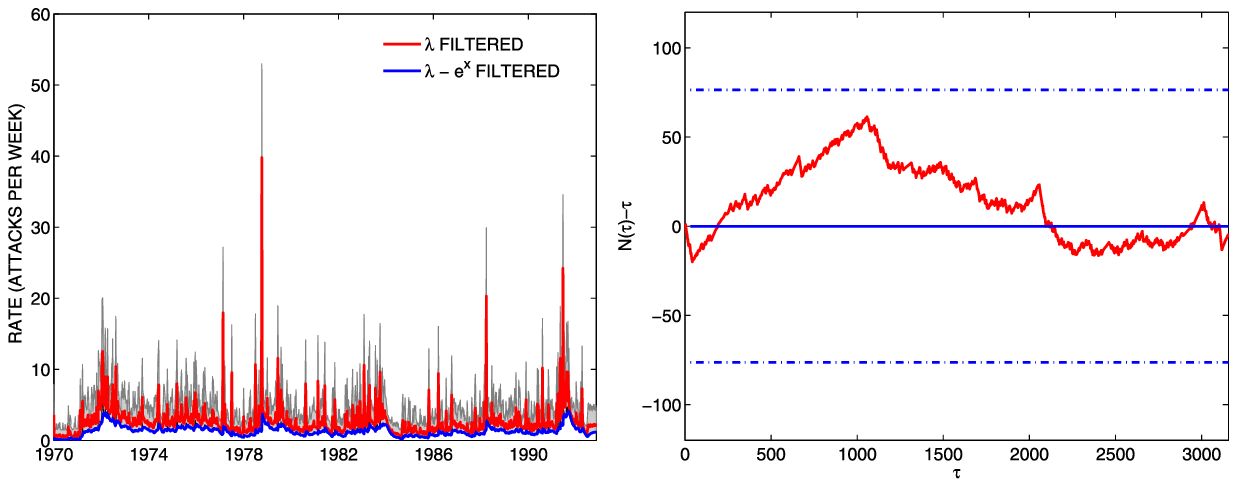}

\caption{Terrorist attacks in Northern Ireland. Left: Filtered
intensity $\lambda$ (posterior mean in red, 95\% range in grey) and
Hawkes rate $\lambda_i -e^{x_i+c\cdot i}$ (posterior mean in blue).
Right: Normalized cumulative distribution of rescaled event times along
with $95\%$ error bounds of the Kolmogorov--Smirnov statistic.}\label{fig8}
\end{figure}

\begin{figure}[b]

\includegraphics{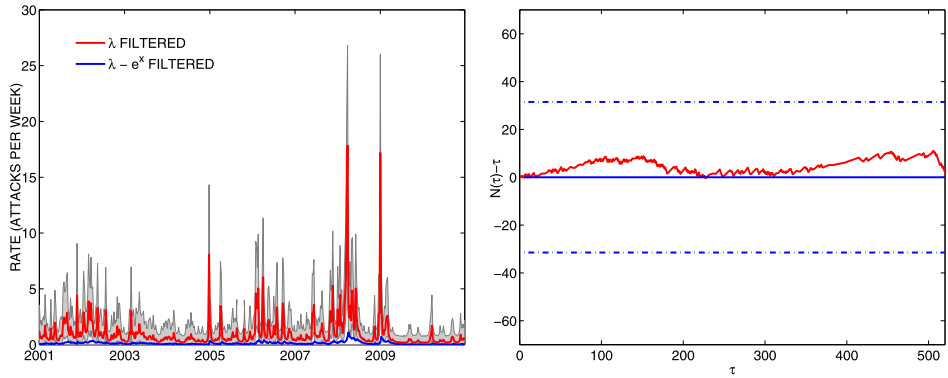}

\caption{Terrorist attacks in Israel. Left: Filtered intensity $\lambda
$ (posterior mean in red, 95\% range in grey) and Hawkes rate $\lambda
-e^x$ (posterior mean in blue). Right: Normalized cumulative
distribution of rescaled event times along with $95\%$ error bounds of
the Kolmogorov--Smirnov statistic.}\label{fig9}
\end{figure}

\begin{figure}

\includegraphics{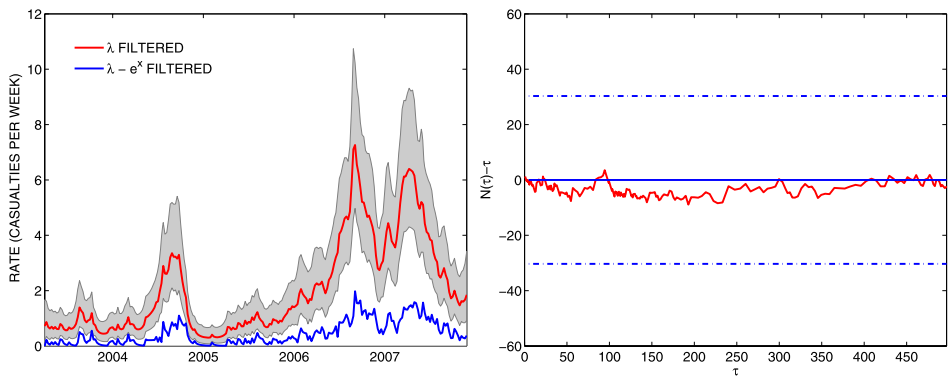}

\caption{Civilian casualties in Fallujah. Left: Filtered intensity
$\lambda$ (posterior mean in red, 95\% range in grey) and Hawkes rate
$\lambda-e^x$ (posterior mean in blue). Right: Normalized cumulative
distribution of rescaled event times along with $95\%$ error bounds of
the Kolmogorov--Smirnov statistic.}\label{fig10}
\end{figure}

In Figure~\ref{fig8} we plot the filtered intensity for terrorist attacks in
Northern Ireland. The estimated dynamics of the process have
similarities to the Chicago crime data set, as the percentage of events
attributed to contagion is $50\%\pm5$\%. The timescale over which the
estimated self-excitation lasts, $9.3$ weeks, is the longest out of all
of the data sets explored here. In contrast, we plot the corresponding
intensities for terrorist attacks in Israel in Figure~\ref{fig9} and for
civilian casualties in Figure~\ref{fig10}. In Israel, we observe very little
contagion effects, similar to those in Figure~\ref{fig5}. The timescale
associated with history independent correlation in Iraq is the slowest
out of all 5 data sets, at 36 weeks. This is likely due to exogenous
factors such as troop surges playing a large role in intensity
fluctuations. However, a significant proportion of clustering is
attributed to contagion effects, verifying previous work in \citet
{lewis2012}. The differences across these three extreme security
settings may be due to a variety of factors, such as the security
measures employed by the government in power, the organization and
tactics of the opposition, the local geography, etc. Trying to
determine theses underlying factors could be an important line of
future research.

\section{Discussion}\label{sec5}

We developed a model and inference methodology to assess the levels of
contagion and correlation in crime and security data. We connected
Hawkes process and Cox process type models that are typically used
independently to explain clustering in crime and security data sets.
The high-dimensional nature of the problem, filtering the latent vector
$\vec{x}$, was overcome by using a model with a sparse covariance
matrix and a Hawkes component that can be written as a differential equation.

Determining whether contagion effects are present in security related
time series is a problem of practical importance. The effectiveness of
policing strategies such as \textit{cops on the dots}, where police react
to recent crimes [\citet{jones2010}], depends on how the crime event
history influences future crime rates. Accurate assessment of the
timescale associated with contagion effects may tell police how long
they need to put additional patrols in a neighborhood. If exogenous
effects are also causing crime rate fluctuations, these effects need to
be taken into account if parameter estimates are to be accurate.
Similar considerations may be relevant to military strategies in
extreme security settings.

For these types of applications, spatial-temporal processes are needed
and we believe our methodology should extend to this setting. In \citet
{girolami2011}, the authors illustrate the feasibility of Hamiltonian
and Langevin Monte Carlo in high-dimensional settings, in particular,
for a 2D LGCP. To add time and self-excitation, several approaches
could be used to model the Cox process: choosing a model with sparse
inverse covariance matrix, modeling the inverse covariance matrix
explicitly, using a sparse approximation, or via a nonparametric sparse
estimator, such as $l_1$ penalization. Spatial extensions will be the
focus of subsequent research.

\printaddresses

\end{document}